\documentclass{PoS}

\newcommand{\psib}{\overline{\psi}}
\newcommand{\phib}{{\overline{\phi}}}
\newcommand{\chib}{{\overline{\chi}}}

\newcommand{\beq}{\begin{equation}}
\newcommand{\eeq}{\end{equation}}
\usepackage{slashed}
\title{Four fermion operators and the search for BSM Physics}

\ShortTitle{Four fermion operators}

\author{\speaker{Simon Catterall}\thanks{Work supported in part by DOE grant DE-FG02-85ER40237}\\
        Physics Department, Syracuse University\\
        E-mail: \email{smc@physics.syr.edu}}

\author{Aarti Veernala\\
        Physics Department, Syracuse University\\
        E-mail: \email{aveernal@syr.edu}}

\abstract{
We report on Monte Carlo simulations focused on elucidating the phase structure of
a $SU(2)$ gauge theory containing $N_f$ Dirac fermion flavors transforming in the fundamental representation of the group and interacting through an additional chirally
invariant four fermion term.  Pairs of physical flavors are
implemented using the two tastes present in a reduced staggered fermion
formulation of the theory with the Yukawa interactions necessary for generating the four
fermion term preserving the usual shift symmetries.
We observe a
crossover in the behavior of the chiral condensate for strong
four fermi coupling associated with the
generation of a dynamical
mass for the fermions. At weak gauge coupling this crossover is consistent with the usual continuous phase
transition seen in the pure (ungauged) NJL model. However, if the gauge coupling is strong enough to cause confinement
we observe a much more rapid crossover in the chiral condensate
consistent with a first order phase transition}

\FullConference{The 30 International Symposium on Lattice Field Theory - Lattice 2012,\\
		June 24-29, 2012\\
		Cairns, Australia}

\begin{document}

\section{Introduction}
Elucidating the nature of the electroweak symmetry
breaking sector of the Standard Model (SM) is the main goal of the Large Hadron Collider currently running at CERN. It
is widely believed that the simplest scenario involving a single scalar Higgs field is untenable due to the fine tuning and triviality problems which arise
in scalar field theories.  One natural solution to these problems can be found by assuming that the Higgs sector
in the Standard Model arises as an effective field theory describing the dynamics of a composite field arising
from strongly bound fermion-antifermion pairs. These models are generically called
technicolor theories.

However, to obtain fermion masses in these scenarios requires additional model building, as in extended technicolor models~\cite{ETC-1,ETC-2,schrock2,schrock3} and models of top-condensation~\cite{Miransky:1988xi,Miransky:1989ds,Bardeen:1989ds,Marciano:1989mj}. In the latter
models four-fermion interactions drive the formation and condensation of a scalar top--anti-top bound state which plays
the role of the Higgs at low energies.

Our motivation in this paper is to study how
the inclusion of such four fermion interactions may influence the
phase structure and low energy behavior of non-abelian gauge theories in general.
Specifically we have examined a model with both gauge interactions and a chirally invariant
four fermi interaction - a model known in the literature as the gauged NJL model \cite{DSB-yamawaki}. 

The focus of the current work is to explore the phase diagram when fermions are charged under a non-abelian gauge group.  Indeed, arguments have been given in the continuum that
the gauged NJL model may exhibit different critical behavior at the boundary
between the symmetric and broken phases \footnote{Notice that the appearance of a true
phase transition in the gauged NJL models depends on the approximation that we can neglect the
running of the gauge coupling} corresponding to the appearance of a line
of new fixed points associated with a mass anomalous dimension varying in the
range
 $1 < \gamma_{\mu} < 2$ \cite{DSB-yamawaki,walking-francesco2}. The evidence for this behavior
derives from calculations utilizing the ladder approximation in Landau gauge to the Schwinger-Dyson equations.  A primary goal of the current study was to use lattice simulation to check the validity of
these conclusions and specifically to search for qualitatively new critical behavior in the
gauged model
as compared to the pure NJL theory. While we will present results that indicate that
the phase structure of the gauged NJL model is indeed different from pure NJL, we shall
argue that our results are \emph{not} consistent
with the presence of any new fixed points in the theory.

In the work reported here and described in detail in \cite{us} we have concentrated on the four flavor theory
corresponding to
two copies of the basic Dirac doublet used in the lattice construction.
The four flavor theory is expected to be chirally
broken and confining at zero four fermi coupling and is free from
sign problems for gauge group $SU(2)$. Understanding the effects of the four fermion term in this
theory can then serve as a benchmark for future studies of theories which, for zero
four fermi coupling, lie near or inside the conformal window. In the latter
case the addition of
a four fermion term will break conformal invariance but in principle that breaking may be made arbitrarily small
by tuning the four fermi coupling. It is entirely possible that the phase diagrams of such conformal or walking
theories in the presence of four fermi terms may exhibit very different features than those seen for
a confining gauge theory.

\section{Details of the model}
We will consider a model which consists of $N_f/2$ doublets of gauged massless Dirac fermions in the
fundamental representation of an $SU(2)$ gauge group and
incorporating an $SU(2)_L\times SU(2)_R$ chirally invariant four fermi interaction.
The action for a single doublet takes the form
\begin{eqnarray}
S &=& \int d^4x\; \psib ( i \slashed{\partial} -  \slashed{A}) \psi - \frac{G^2}{2N_f} [ (\bar{\psi} \psi)^{2} + (\bar{\psi} i \gamma_{5} \tau^{a} \psi )^{2} ]  \nonumber \\
&-& \frac{1}{2g^2} Tr [F_{\mu \nu} F^{\mu \nu}] ,
\label{eq:etcnjlaction}
\end{eqnarray}
where G is the four-fermi coupling, $g$ the usual gauge coupling
and $\tau^{a},a=1\ldots 3$ are the generators of the $SU(2)$ flavour group.

This action may be discretized using the (reduced) staggered fermion formalism with the result
\beq
S= \sum_{x,\mu} \ \chi^{T}(x) \ \mathcal{U}_{\mu}(x) \ \chi(x + a_{\mu}) \ [\eta_{\mu}(x) +G\; \phib_\mu(x) \,\epsilon(x) \, \xi_\mu(x)] .
\label{finalS-latt} \eeq where $\eta_\mu(x)$, $\xi_\mu(x)$ and $\epsilon(x)$ are the usual
staggered fermion phases, $\phib(x)=\frac{1}{16}\sum_h \phi(x-h)$ the average of the scalar field over the hypercube \cite{redstag-Smit-1, redstag-Smit-2} and the gauge field acting on the reduced staggered fermions takes the form:
\beq \mathcal{U}_{\mu} (x) = \frac{1}{2} [1+ \epsilon(x)] \; U_{\mu}(x) + \frac{1}{2} [1- \epsilon(x)] \; U_{\mu}^{*}(x) \label{mathcalU}. \eeq

Clearly the theory is invariant under the $U(1)$
symmetry $\chi(x)\to e^{i\alpha\epsilon(x)}\chi(x)$ which is to be interpreted as the $U(1)$ symmetry corresponding to
fermion number.
More interestingly it is also invariant under
certain shift symmetries given by 
\begin{eqnarray}
\chi(x)&\to&\xi_\rho(x)\chi(x+\rho) , \\
U_\mu(x)&\to&U_\mu^{*}(x+\rho) , \\
\phi_\mu(x)&\to&(-1)^{\delta_{\mu\rho}}\phi_\mu(x+\rho) .
\end{eqnarray}
These shift symmetries
correspond to a {\it discrete} subgroup of
the continuum axial flavor transformations which act on the matrix field $\Psi$ according to
\beq \Psi\to \gamma_5\Psi\gamma_\rho\eeq
Notice that no single site mass term is allowed in this model.

\section{Numerical results}

We have used the RHMC algorithm to simulate the lattice theory  with
a standard Wilson gauge action being employed for the gauge fields. Upon integration over
the basic fermion doublet we obtain a Pfaffian ${\rm Pf(M(U))}$ depending on the gauge field \footnote{Note that the fermion operator appearing in eqn.~\ref{finalS-latt} is antisymmetric}.
The required pseudofermion weight for $N_f$ flavors is then
${\rm Pf}(M)^{N_f/2}$. The pseudoreal character of
$SU(2)$ allows us to show that the
Pfaffian is purely real and so we are guaranteed to have no sign problem if
we use multiples of four flavors corresponding to a
pseudofermion operator of the form  $(M^\dagger M)^{-{\frac{N_f}{8}}}$. The results in this
paper are devoted to the case $N_f=4$.
We have utilized
a variety of lattice sizes: $4^4$, $6^4$, $8^4$ and
$8^3\times 16$ and a range of gauge couplings $1.8< \beta \equiv 4/g^2< 10.0$.
\begin{figure}[htb]
\begin{center}
\includegraphics[height=70mm]{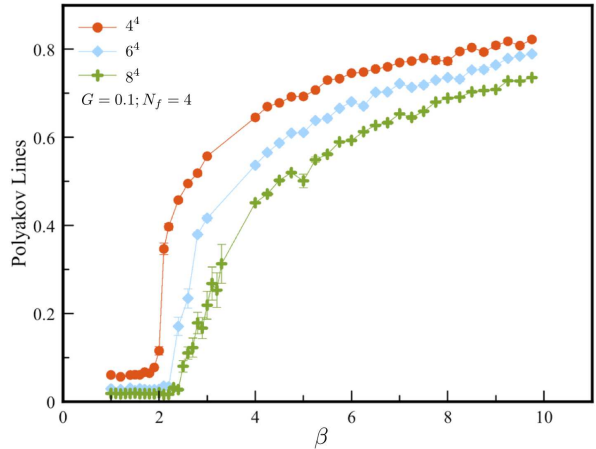}
\caption{Polyakov loop vs $\beta$ at $G=0.1$ for four flavours}
\label{poly-L468}
\end{center}
\end{figure}
To determine where the pure gauge theory is strongly coupled and confining we
have examined the average Polyakov line
as $\beta$ varies holding
the four fermi coupling fixed at $G=0.1$.
This is shown in figure ~\ref{poly-L468}.
We see a strong crossover between a confining regime for small $\beta$
to a deconfined regime at large $\beta$. The crossover coupling is volume dependent and takes
the value of $\beta_c\sim 2.4$  for lattices of size $L=8$.
For $\beta<1.8$ the plaquette
drops below 0.5 which we take as indicative of the presence of strong lattice spacing artifacts and so
we have confined our simulations to larger values of $\beta$.
We have set the fermion mass to zero in all of our work so that our lattice
action possesses the series of exact chiral symmetries discussed earlier.

One of the primary observables used in this
study is the chiral condensate which
is computed from the gauge invariant one link mass operator
\beq
\chi (x)\left({\mathcal U}_\mu(x)\chi (x+e_\mu)+{\mathcal U}^\dagger_\mu(x-e_\mu)\chi(x-e_\mu)\right) \epsilon (x)\xi_\mu(x)\eeq

\begin{figure}
\begin{center}
\includegraphics[height=70mm]{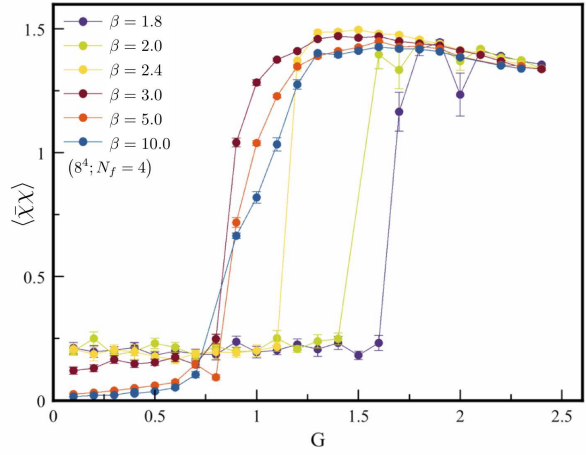}
\caption{$\langle\chib\chi\rangle$ vs $G$ for varying $\beta$ for the $8^4$ lattice with $N_f = 4$.}
\label{oct12-psibpsi-L8-N4-all-beta}
\end{center}
\end{figure}
In Figure \ref{oct12-psibpsi-L8-N4-all-beta} we 
show a plot of the absolute value of the condensate at a variety of gauge couplings
$\beta$ on $8^4$ lattices.  Notice the rather smooth transition between symmetric and broken phases around
$G\sim 0.9$ for $\beta = 10$. This is consistent with earlier work using sixteen flavors of naive fermion reported
in \cite{annakuti} which identified a line of second order phase transitions in this region of
parameter space. It also agrees with the behavior seen in previous simulations using conventional staggered quarks \cite{Hands:1997uf}.

\begin{figure}
\begin{center}
\includegraphics[height=70mm]{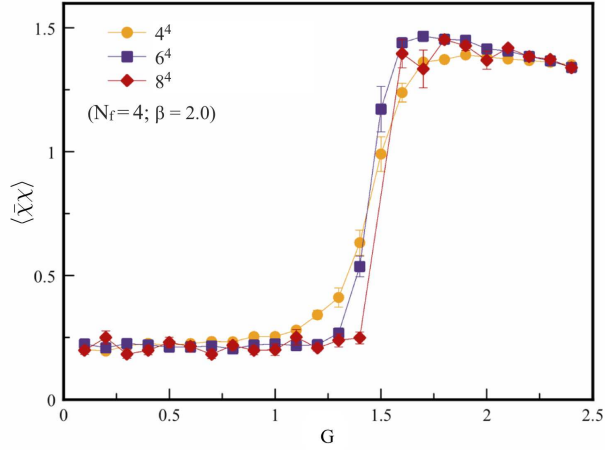}
\caption{$\langle\chib\chi\rangle$ vs $G$ at $\beta=2.0$ for lattices $4^4$, $6^4$ and $8^4$ with $N_f = 4$.}
\label{oct12-psibpsi-all-L-N4}
\end{center}
\end{figure}

This behavior should be contrasted with the behavior of the condensate for strong gauge coupling $\beta\le 2.4$. Here a very
sharp transition can be seen reminiscent of a first order phase transition. In Figure~\ref{oct12-psibpsi-all-L-N4} we highlight this
by showing a plot of the condensate versus four fermi coupling at the single gauge coupling $\beta=2.0$ for a range of
different lattice sizes. 
The chiral condensate is now non-zero even for small four fermi coupling and shows no
strong dependence on the volume consistent with spontaneous chiral symmetry breaking in the pure gauge
theory. However, it jumps abruptly to much
larger values when the four fermi coupling exceeds some critical value.
This crossover or transition
is markedly discontinuous in character - reminiscent of a first order phase transition. Indeed,
while the position of the phase transition is only weakly volume dependent it appears
to get sharper with increasing volume.

What seems clear is that the second order transition seen in the
pure NJL model is no longer present when the gauge coupling is strong.
In the next section we will argue that this is to be expected -- in the gauged
model one can no longer send the fermion mass to zero by adjusting the four fermi coupling since it receives
a contribution from gauge mediated chiral symmetry breaking. Indeed the measured one link chiral condensate operator is not an order parameter
for such a transition since we observe it to be non-zero for all $G$. Notice however that  we see no sign
that this condensate depends on the gauge coupling $\beta$ in the confining regime at small $G$. This is qualitatively different from the behavior  of regular staggered
quarks and we attribute it to the fact that the reduced formalism does not allow for a single site mass term or an
exact {\it continuous} chiral symmetry. Thus the spontaneous breaking of the
residual discrete lattice chiral symmetry by gauge interactions will
not be signaled by a light Goldstone pion and the measured condensate will receive contributions only from massive states.
The transition we observe  is probably best thought
of as a crossover phenomenon corresponding to the sudden onset of a new mechanism for dynamical mass generation due to the strong four fermi
interactions.

\section{Summary}

In this paper we have conducted numerical simulations
of the gauged NJL model for four flavors of Dirac fermion in the
fundamental representation of the $SU(2)$ gauge group.
We have employed a reduced staggered fermion discretization scheme which allows us
to maintain an exact subgroup of the continuum
chiral symmetries.

We have examined the model for a variety of values for lattices size, gauge coupling, and four fermi interaction strength. In the NJL limit $\beta\to\infty$
we find evidence for a continuous phase transition for $G\sim 1$ corresponding to
the expected spontaneous breaking of chiral symmetry. However, for gauge couplings that
generate a non-zero chiral condensate even for $G=0$ this transition or crossover appears
much sharper and  there is no evidence of critical fluctuations in the
chiral condensate.

Thus our results are consistent  with the idea that the second order phase transition which exists in the pure
NJL theory ($\beta=\infty$) survives at weak gauge coupling. However our results indicate that
any continuous transition ends if the gauge coupling becomes strong enough to cause confinement.
In this case we do however see evidence of additional dynamical mass generation
for sufficiently large four fermi coupling associated with
an observed rapid crossover in the chiral condensate and a possible first
order phase transition.

The fact that we find the condensate non-zero and constant for strong
gauge coupling and $G< G_{c}$ shows that the chiral symmetry of the theory is already broken as expected for $SU(2)$ with $N_f=4$ flavors.  This
breaking of chiral symmetry due to the gauge interactions is accompanied by the
generation of a non-zero fermion mass even for small four fermi coupling. Notice that
this type of scenario is actually true of top quark condensate models in which the
strong QCD interactions are already expected to break chiral symmetry
independent of a four fermion top quark operator.
The magnitude of
this residual fermion mass is {\it not} controlled by the four fermi coupling and cannot
to sent to zero by tuning the four fermi coupling - there can be no continuous phase
transition in the system as we increase the four fermi coupling - rather the condensate
becomes strongly enhanced for large $G$.

\begin{acknowledgments}
The simulations were carried out using  USQCD
resources at Fermilab and Jlab.
\end{acknowledgments}



\end{document}